\documentclass{aastex}
\usepackage{spr-astr-addons}
\usepackage{url}
\usepackage{amsmath}
\usepackage{graphicx}
\usepackage{lscape}

\RequirePackage{color}
\newcommand{\MC}{\multicolumn}

\DeclareRobustCommand{\ion}[2]{%
\relax\ifmmode
\ifx\testbx\f
{\mathrm{#1\,\textsc{#2}}}\else
{\mathrm{#1\,\mathsc{#2}}}\fi
\else\textup{#1\,{\mdseries\textsc{#2}}}%
\fi}
\def\fbs{\textsc{fbs}}
\newcommand{\parsec}{\texttt{PARSEC}}
\newcommand{\basti}{\texttt{BaSTI}}
\newcommand{\mist}{\texttt{MIST}}

\begin{document}

\title{Long-period eclipsing binaries: towards the true mass-luminosity relation.
    II. Absolute parameters of the NN\,Del system}
%\footnotemark[0]\thanks{%
%This work is based on observations obtained with the Southern African Large Telescope (SALT)
%under programs  \mbox{2017-1-MLT-001} and \mbox{2019-1-SCI-004} (PI: Kniazev).}}

%% Running heads
\shorttitle{The long-period eclipsing binary star NN\,Del}
\shortauthors{Alexei Kniazev}

\author{Alexei Kniazev\altaffilmark{1,2,3}}
\email{akniazev@saao.ac.za}

\altaffiltext{1}{South African Astronomical Observatory, PO Box 9, 7935 Observatory, Cape Town,South Africa.}
\altaffiltext{2}{Southern African Large Telescope Foundation, PO Box 9, 7935 Observatory, Cape Town, South Africa.}
\altaffiltext{3}{Sternberg Astronomical Institute, Lomonosov Moscow State University, Moscow, Russia.}

\begin{abstract}
We present results of our study of the long-period eclipsing binary star NN\,Delphini 
(hereafter NN\,Del). The results are based on spectral data obtained 
with the HRS \'echelle spectrograph of the Southern African Large Telescope (SALT). 
Our constructed velocity curve is based on 19 spectra obtained between 2017 and 2019 years
and covers all phases of the binary's orbit. 
The orbital period, $P=99.252$~days, was determined from our spectral data and
coincides with the period determined in previous studies, as well as the system
eccentricity of $e=0.517$. Calculated velocity amplitudes of both components
allow us to determine the masses of both system components 
$M_1 = 1.320~M_\odot$ and $M_2 = 1.433~M_\odot$ with the accuracy about of 
one percent (0.8\% and 1.1\%), respectively.
Luminosities of both components are presented
as $L_1 = 4.164~L_\odot$ and $L_2 = 6.221~L_\odot$,
and the effective temperatures of both components were directly evaluated
($T_\mathrm{eff1} = 6545$~K and $T_\mathrm{eff2} = 6190$~K)
together with the metallicity of the system $\mathrm{[Fe/H]} = -0.19$~dex
and its color excess E(B-V)=0.026~mag.
Comparison with evolutionary tracks shows that the system age is $2.25\pm0.19$~Gyr, 
and both components are on the main sequence and have not yet passed the turn point.
Spectral type is F5V for the hotter component and F8V for another one.
\end{abstract}

\keywords{stars: luminosity function, mass function --- stars: binaries: spectroscopic --- stars: individual (NN Del)}

%-----------------------------------------------------------------------------------------
\begin{table*}[th]
   \small
   \centering
   \caption{Parameters of NN\,Del system collected from the literature.}
   \begin{tabular}{lllll}
      \hline\hline
		Parameter                                    &   GF (2003)$^a$        &    G(2014)$^b$     &   S(2018)$^c$      &    G(2019)$^d$          \\ \hline
Orbital period $P$ (d)                               & 99.2684$\pm$0.0005     &  99.244$\pm$0.019  & 99.26849$\pm$0.00015 & 99.2690$\pm$0.0009    \\
Eccentricity $e$                                     & 0.51759$\pm$0.00002    &  0.5168$\pm$0.0029 & 0.51944$\pm$0.00055  & 0.5197$\pm$0.0004     \\
Rad.\ vel.\ semi-amplitude $K1$ (km s$^{-1}$)        & --                     &  39.45$\pm$0.27    & 39.62$\pm$0.15       & 39.407$\pm$0.037      \\
Rad.\ vel.\ semi-amplitude $K2$ (km s$^{-1}$)        & --                     &  36.04$\pm$0.21    & 36.22$\pm$0.11       & 36.191$\pm$0.023      \\
Systemic heliocentric vel.\ $\gamma$ (km s$^{-1}$)   & --                     &  $-$8.19$\pm$0.08  & $-$9.405$\pm$0.050   & $-$9.485$\pm$0.015    \\
Inclination $i$ (degrees)                            & 89.488$\pm$0.003       &  90.0 (fixed)      & 89.6342$\pm$0.0076   & 89.90$\pm$0.11        \\
The longitude of the periastron $\omega$  (degrees)  & 171.710$\pm$0.005      &  170.0$\pm$0.4     & 169.75$\pm$0.45      & 170.00$\pm$0.07       \\
R.m.s. residuals of Keplerian fit (km s$^{-1}$)      &   --                   &  0.56              & 0.11                 & 0.10 (0.18)           \\
$m_1$ (M$_{\odot}$) (Component A)                    &   --                   &  1.328$\pm$0.021   & 1.337$\pm$0.011      & 1.3266$\pm$0.0021     \\
$m_2$ (M$_{\odot}$) (Component B)                    &   --                   &  1.454$\pm$0.025   & 1.462$\pm$0.013      & 1.4445$\pm$0.0029     \\
$r_1$                                                & 0.01584$\pm$0.00006    &  --                & 0.01260$\pm$0.00015  & --                    \\
$r_2$                                                & 0.0153                 &  --                & 0.01720$\pm$0.00015  & --                    \\
$T_\mathrm{eff}$ (K) (Component A)                   & 6437$\pm$5             &  --                & --                   & --                    \\
$T_\mathrm{eff}$ (K) (Component B)                   & 6500.0                 &  --                & --                   & --                    \\
$\log g$ (dex) (Component A)                         &   --                   &  --                & 4.1532$\pm$0.0051    & --                    \\
$\log g$ (dex) (Component B)                         &   --                   &  --                & 3.9226$\pm$0.0035    & --                    \\
Radius R (R$_{\odot}$)  (Component A)                &   2.2                  &  --                & 1.604$\pm$0.014      & --                    \\
Radius R (R$_{\odot}$)  (Component B)                &   2.2                  &  --                & 2.188$\pm$0.015      & --                    \\
$\mathrm{v \sin i}$  (km s$^{-1}$) (Component A)     &   --                   &  7.7$\pm$0.6       & --                   & --                    \\
$\mathrm{v \sin i}$  (km s$^{-1}$) (Component B)     &   --                   &  7.4$\pm$0.3       & --                   & --                    \\
Spectral Type                                        &   F5 IV (both)         &  F8 IV-V (both)    & --                   & --                    \\
\hline
\MC{5}{l}{$^a$ -- \citet{2003Ap&SS.283..297G}; $^b$ -- \citet{2014Obs...134..109G}; $^c$ -- \citet{2018MNRAS.478.1942S}; $^d$ -- \citet{2019A&A...632A..31G} }
   \end{tabular}
   \label{tab:NN_Del_publ}
\end{table*}
%E(B-V) (mag)                                         &   0.1                  &  --                & --                   & --                    \\
%Epoch at radial velocity maximum $T_0$ (d)           & 2450227.6026$\pm$0.0015&  2456389.39$\pm$0.06 \\
%Metallicity (dex)                                    &   --                   &  --                  \\

%-----------------------------------------------------------------------------------------

%**************************************************************************
\section{Introduction}

The ratio between the mass of a star and its luminosity on the main sequence 
(mass-luminosity ratio, MLR hereafter) is a fundamental law used in various fields of astrophysics.
It is particularly important to restore the initial mass function (IMF) using
the luminosity function (LF) of stars, 
and it should be noted that the IMF can only be obtained by this method, that is,
through the LF and MLR.

Independent determination of the mass of a star and its luminosity is only possible 
for binary system components of certain types. 
One of these types are visual binary stars with known orbital parameters and trigonometric parallaxes
\citep[][]{1998A&A...338..455F,2012A&A...546A..69M,2016MNRAS.459.1580D}.
Usually, these stars are wide pairs, whose components do not interact with each other
and, in an evolutionary sense, are similar to single stars.
Another main source of independent mass determinations are detached eclipsing binary stars 
with components on the main sequence, whose spectra show lines of both components
\citep[double-lined eclipsing binaries, DLEB;][]{1980ARA&A..18..115P,1991A&ARv...3...91A,1998ARep...42..793G,2001ARep...45..972K,2010A&ARv..18...67T}.
These are mainly close pairs whose component's spins are synchronized by tidal interaction, 
and they evolve slightly different compare to single stars due to the slowdown of rotation. 
At the same time, for masses M/M$_{\odot} > 2.7$, MLR is based exclusively
on the data obtained for DLEBs, and it is used for studies of single stars.
Apparently, such use of MLR could be not very correct and leads to systematically wrong results 
when evaluating the characteristics of stars in the discussed mass range, 
as well as when restoring the initial mass function with its help. 
For example, \citet{2003A&A...402.1055M} compared the radii of DLEBs and single stars 
and found a noticeable difference between the observation parameters of the B0V-G0V components 
of the DLEBs and single stars of similar spectral classes.

\citet{2007MNRAS.382.1073M} collected data on fundamental parameters of 19 components 
of long-period DLEB. These stars presumably have not undergone synchronization 
of rotation with the orbital period and therefore rotate rapidly
and evolve similarly to single stars. 
Possibly, only such kind of data should be used to construct relations 
(in particular, MLR) for ``isolated'' stars for masses M/M$_{\odot} > 2.7$.
The masses of the components of other types of binary stars (orbital, resolved spectral binaries) 
rarely exceed this limit. Note that of the 19 DLEB stars mentioned above, 
only 13 have a mass greater than 2.7 M/M$_{\odot}$, which is clearly insufficient 
to draw any definite conclusions. 
For this reason, we have begun a systematic study of long-period DLEB stars 
to obtain their mass and luminosity parameters and to compare these parameters
with those obtained for short-period DLEB stars.
The definition of the test sample as well as a brief description of the software package \fbs\
created to analyze spectra of binary stars is described in our first paper \citep{2020arXiv200404115K}.
Here, we present the first results obtained for the long-period eclipsing
binary star NN\,Del, which belongs to our test sample and has some previously
published studies to help evaluate the accuracy of the parameters based on our spectral data. 
Hereafter we call the brighter star in the DLEB system NN\,Del component A.

%--------------------------------------------------------------------------
\begin{table*}
\centering
\caption{HRS observations for NN\,Del and found heliocentric velocities}
\label{tab:NN_Del_obs}
\begin{tabular}{cccrr}
\hline\hline
Date      &   MJD         &  Exp.time  &\MC{1}{c}{V$_{hel1}$}&\MC{1}{c}{V$_{hel2}$}\\
	  &   (day)       &  (Sec)     &\MC{1}{c}{(km/sec)}  &\MC{1}{c}{(km/sec)}  \\
  (1)     &   (2)         &  (3)       &\MC{1}{c}{(4)}       &\MC{1}{c}{(5)}       \\
\hline
20170504  & 2457878.65465 &  315.      &  43.528$\pm$0.012   &  -68.014$\pm$0.025  \\
20170508  & 2457882.63785 &  315.      &  38.422$\pm$0.012   &  -61.257$\pm$0.026  \\
20170520  & 2457894.64881 &  400.      &  -2.161$\pm$0.045   &  -16.526$\pm$0.041  \\
20170805  & 2457971.46249 &  315.      &  12.869$\pm$0.013   &  -34.007$\pm$0.034  \\
20170814  & 2457980.36746 &  450.      &  42.930$\pm$0.016   &  -66.858$\pm$0.029  \\
20170830  & 2457996.37722 &  315.      &  -7.087$\pm$0.023   &  -12.431$\pm$0.039  \\
20170927  & 2458024.28117 &  315.      & -25.935$\pm$0.017   &    8.459$\pm$0.021  \\
20171003  & 2458030.27510 &  315.      & -26.714$\pm$0.017   &    9.568$\pm$0.024  \\
20171024  & 2458051.23722 &  240.      & -23.874$\pm$0.019   &    5.794$\pm$0.035  \\
20180428  & 2458237.66367 &  315.      & -28.411$\pm$0.029   &    8.636$\pm$0.056  \\
20180429  & 2458238.65980 &  315.      & -27.308$\pm$0.014   &    8.903$\pm$0.031  \\
20180926  & 2458388.33887 &  315.      &   3.222$\pm$0.014   &  -23.580$\pm$0.024  \\
20180930  & 2458392.23932 &  315.      &  -5.554$\pm$0.016   &  -15.491$\pm$0.031  \\
20181007  & 2458399.29359 &  315.      & -14.738$\pm$0.016   &   -5.158$\pm$0.032  \\
20181015  & 2458407.27283 &  315.      & -21.292$\pm$0.015   &    2.574$\pm$0.027  \\
20190728  & 2458693.50237 &  315.      & -10.364$\pm$0.041   &   -8.896$\pm$0.070  \\
20190801  & 2458697.46381 &  315.      & -14.558$\pm$0.016   &   -4.101$\pm$0.031  \\
20190810  & 2458706.45261 &  315.      & -21.766$\pm$0.016   &    3.275$\pm$0.030  \\
20190818  & 2458714.44764 &  315.      & -24.539$\pm$0.014   &    7.221$\pm$0.030  \\
\hline\hline
\end{tabular}
\end{table*}
%--------------------------------------------------------------------------

%**************************************************************************
\section{Previous studies of NN\,Del}
\label{txt:NN_Del}

The system NN\,Del (HD\,197952, HIP\,102545) was discovered as variable based on {\sc Hipparcos}
satellite data \citep{1994IBVS.4118....1M}, but the only information 
published was the star changing its brightness by more than half of a magnitude.
The brightness curve for the NN\,Del system was obtained and thoroughly 
studied by \citet{2003Ap&SS.283..297G}. They concluded that both components of the system 
are almost equal to each other, the period of the system is 99.268~days 
and the eccentricity of the system is 0.5176, i.e. the stellar components have non-circular orbits.
\citet{2003Ap&SS.283..297G} also determined the types of each star as F5,
based on photometric data, and since the radii of both components were calculated 
to be 2.2~R$_{\odot}$, which meant that both stars had already left the main sequence,
their luminosity class was defined as IV.
Table~\ref{tab:NN_Del_publ} shows all parameters and errors, for NN\,Del
that were defined by \citet{2003Ap&SS.283..297G} and following works,
where this binary system was studied.

\citet{2014Obs...134..109G} carried out spectroscopy of NN\,Del using 
a photoelectric spectrometer designed to measure the radial velocities of stars 
\citep{1967ApJ...148..465G}. It should be noted here that the spectra were not obtained, 
but the velocities of both components were determined. 
In total, 37 measurements were made during three years, and for the first time a velocity curve 
was constructed and masses of both components were determined. 
The final accuracy of the velocity curve (scattering of residuals after model subtraction) 
was 0.56~km~sec$^{-1}$. The obtained values of the orbital period $P$ and eccentricity $e$ 
agreed with numbers found from photometric data \citep{2003Ap&SS.283..297G}, 
where the inclination $i$ of the system was assumed to be 90~degrees, but the longitude $\omega$ 
of the periastron $w$ was different at the 4$\sigma$ level, although \citet{2014Obs...134..109G} 
noted the unbelievable high accuracy of all the parameters given by \citet{2003Ap&SS.283..297G}.

\citet{2018MNRAS.478.1942S} carried out spectral observations of NN\,Del 
using the \'echelle spectrograph CHIRON (R~$\sim 80000$) installed on the 1.5m telescope of CTIO
observatory (Chile). Only seven observations were taken, but the spectroscopy was combined 
with photometry from \citet{2003Ap&SS.283..297G} to construct both the velocity
and the brightness curves simultaneously. The obtained masses of both components of the system 
agreed with those obtained by \citet{2014Obs...134..109G} taking into account errors.
The accuracy of the velocity curve was 0.11~km~sec$^{-1}$ that is much higher compare
to \citet{2014Obs...134..109G}.
The use of spectral and photometric data together made it possible to determine masses 
of both components and their radii, although the authors noted that defining all 
these characteristics was not the main purpose of their work.
For the same reason, the authors did not try to estimate 
temperatures and luminosities of both components from the obtained spectra,
and presented only the ratio of the luminosities of the components as 
L$_{\rm B}$/L$_{\rm A} = 1.6564\pm0.0076$.

%========================================================================
\begin{figure}[t]
\centering{
 \includegraphics[clip=,angle=0,width=8.0cm]{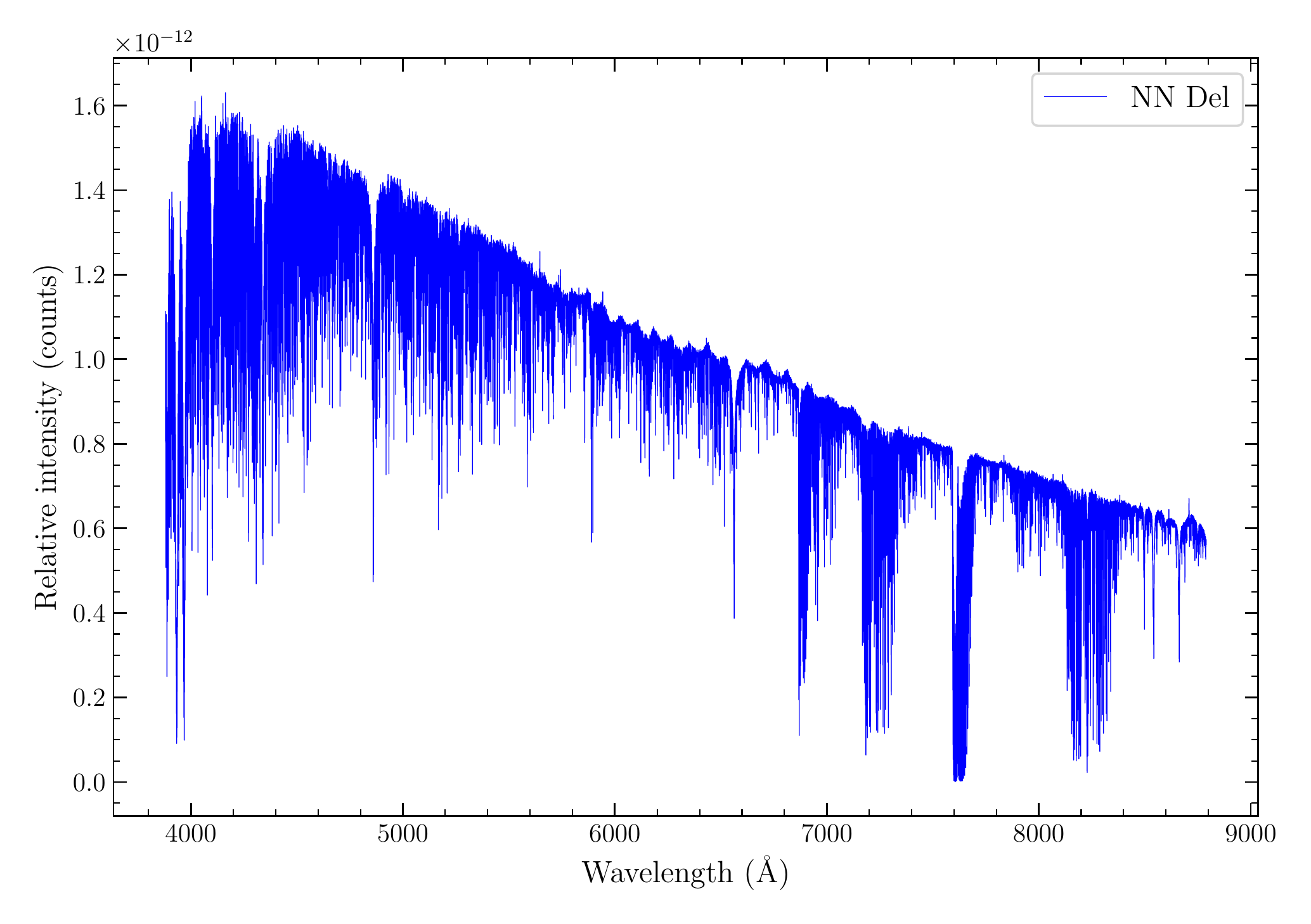}
}
 \caption{An example of a fully processed spectrum of NN\,Del.
The spectrum consists of 70 \'echelle orders from both blue and red arms
merged together and corrected for sensitivity.
 \label{fig:NN_Del_spec}}
\end{figure}
%========================================================================

\citet{2019A&A...632A..31G} is the most recent work in which the characteristics 
of the NN\,Del system  were defined. 
In that paper spectral data were obtained with the \'echelle spectrograph STELLA (R~$\sim 55000$) 
and were used along with interferometric data obtained with VLTI/PIONIER.
40 \'echelle spectra were taken during 2014--2018. 
The simultaneous use of spectral and astrometric data allowed the determination of the masses
of both components and the distance to NN\,Del.
The accuracy of the velocity curve was 0.10~km~sec$^{-1}$ for component A
and 0.18~km~sec$^{-1}$ for component B, which made it possible to determine the masses 
of the components with an accuracy of 0.18\%, and the distance to the system NN\,Del 
with an accuracy of 0.4\%. It was noted that the resulting parallax is greater
than the parallax in $Gaia$ DR2 \citep{2018A&A...616A..11G},
but agrees with the estimate obtained by {\sc Hipparcos}.
Further, \citet{2019A&A...632A..31G} tried to evaluate
the evolutionary status of both components of NN\,Del.
They used their own mass estimates of both components, obtained with high 
accuracy and adopted temperatures, fractional radii and luminosities 
for both components of NN\,Del.
Finally, after analysis of all available data, it was concluded that both components 
of the NN\,Del system are on the main sequence and have not yet passed the turn point.

%**************************************************************************
\section{Spectral observations and data reduction}
\label{txt:Obs_and_Red}

Spectral observations of NN\,Del were made during 2017--2019 at the Southern African Large Telescope 
\citep[SALT;][] {2006SPIE.6267E..0ZB,2006MNRAS.372..151O} 
using the fiber \'echelle spectrograph HRS
\citep[][]{2008SPIE.7014E..0KB,2010SPIE.7735E..4FB,2012SPIE.8446E..0AB,2014SPIE.9147E..6TC}. 
HRS is a thermostabilised dual-beam \'echelle spectrograph. 
The blue arm of the spectrograph covers the spectral range 3900--5550~\AA, 
and the red arm covers the spectral range 5550--8900~\AA, respectively. 
HRS is equipped with four pairs of fibers (object fiber and sky fiber) 
and can be used in low (LR), medium (MR), high resolution (HR) and high stability modes. 
All spectral observations of NN\,Del were taken with MR mode (R=36\,500--39\,000),
with fibers with a diameter of 2.23 arcseconds. 
Both the blue arm and the red arm CCDs were used with binning 1$\times$1.
The dates of observations, their Julian days, and exposure times are presented 
in Table~\ref{tab:NN_Del_obs}.

Since HRS is the vacuum \'echelle spectrograph that is installed inside a temperature-controlled
enclosure, all standard calibrations are performed once a week, 
which is enough to achieve an accuracy 300~m/sec for MR mode. 
The primary HRS data reduction was done automatically using the standard SALT pipeline described 
by \citet{2010SPIE.7737E..25C}. The following \'echelle data reduction was performed using the 
standard HRS data pipeline described in detail by \citet{2016MNRAS.459.3068K,2019AstBu..74..208K}.

Each HRS spectrum of NN\,Del was additionally corrected for bad columns and pixels 
and was also corrected for the spectral sensitivity curve obtained closest to the date 
of the observation. Spectrophotometric standards for HRS are observed once a week 
as part of the HRS Calibration Plan.
An example of a fully processed spectrum of NN\,Del, which was used in further analysis,
is shown in Figure~\ref{fig:NN_Del_spec}. 
The spectrum consists of 70 \'echelle orders from both the blue and the red arm 
merged together and corrected for sensitivity.
Unfortunately, SALT is a telescope where the unfilled entrance pupil of the telescope
moves during the observation. For that reason, absolute flux calibration is not
feasible with SALT. However, since all optical elements are always the same, 
relative flux calibration can be used for SALT data.
Additionally, since HRS is a fiber-fed \'echelle spectrograph,
the relative distribution of energy could be done with high enough accuracy
(Kniazev 2020; in preparation).

%**************************************************************************
\section{Available photometric data}
\label{txt:Photometry}

%========================================================================
\begin{figure}[t]
\centering{
 \includegraphics[clip=,angle=0,width=8.0cm]{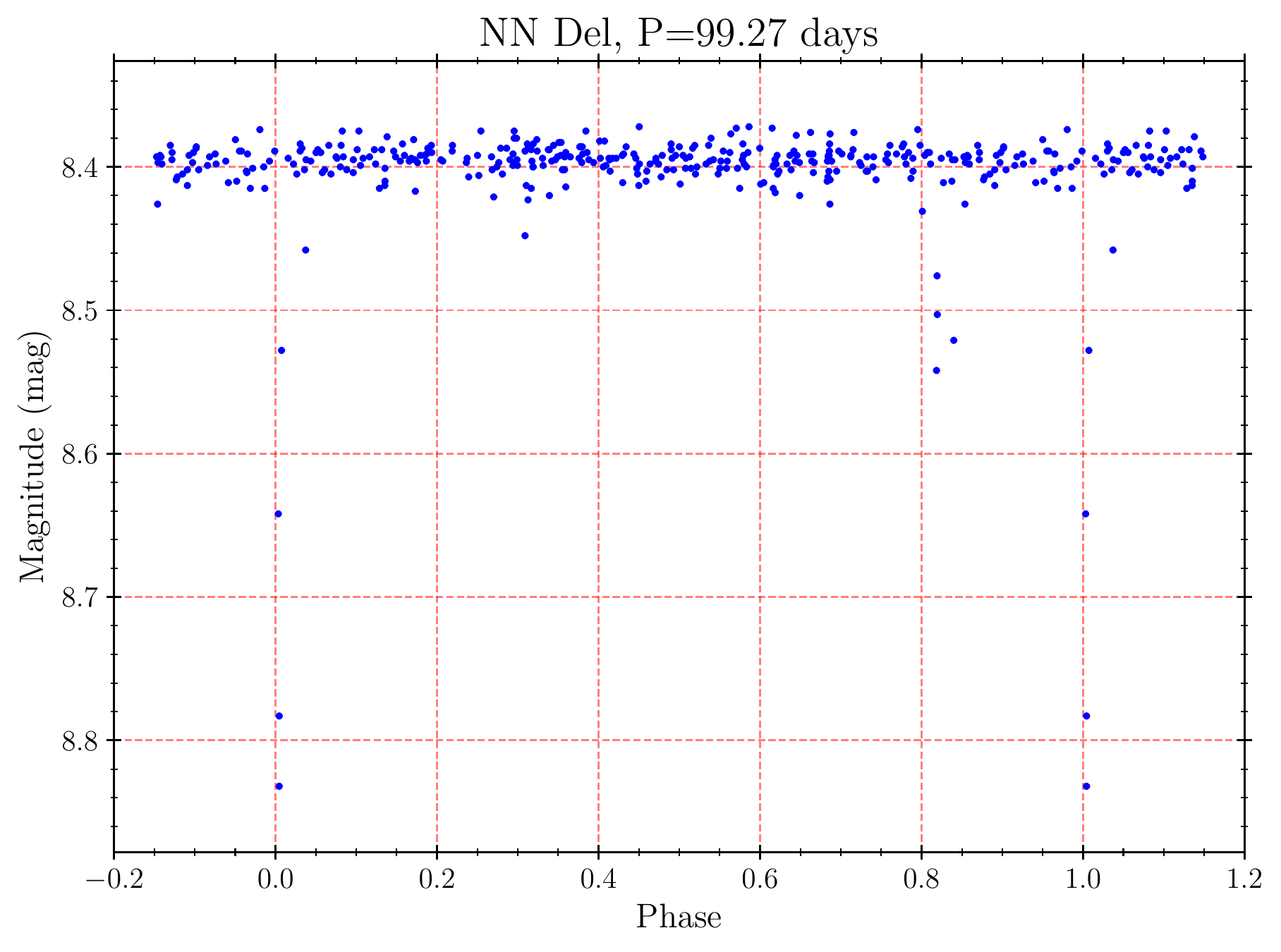}
}
 \caption{Photometric data from the ASAS survey converted to the period P=99.27~days. 
     There are only few points that can indicate the shape of the primary and secondary minima. 
 \label{fig:NN_Del_phot}}
\end{figure}
%========================================================================

Unfortunately, there are no good photometric data for NN\,Del among all existing public surveys.
The best available data are from the ASAS survey \citep{1997AcA....47..467P} as shown 
in Figure~\ref{fig:NN_Del_phot}. It is obvious, that even these data 
have too few points outlining the positions and shapes of the narrow primary 
and secondary minima and it is impossible to use these data for any modeling. 
Therefore, it was decided to use our spectral data together with the results of the analysis
of photometric data from \citet{2003Ap&SS.283..297G}, 
which were also used by \citet{2018MNRAS.478.1942S}.

%**************************************************************************
\section{Spectral data analysis}
\label{txt:spec_analysis}

%========================================================================
\begin{figure*}
\centering{
 \includegraphics[clip=,angle=0,width=12.9cm]{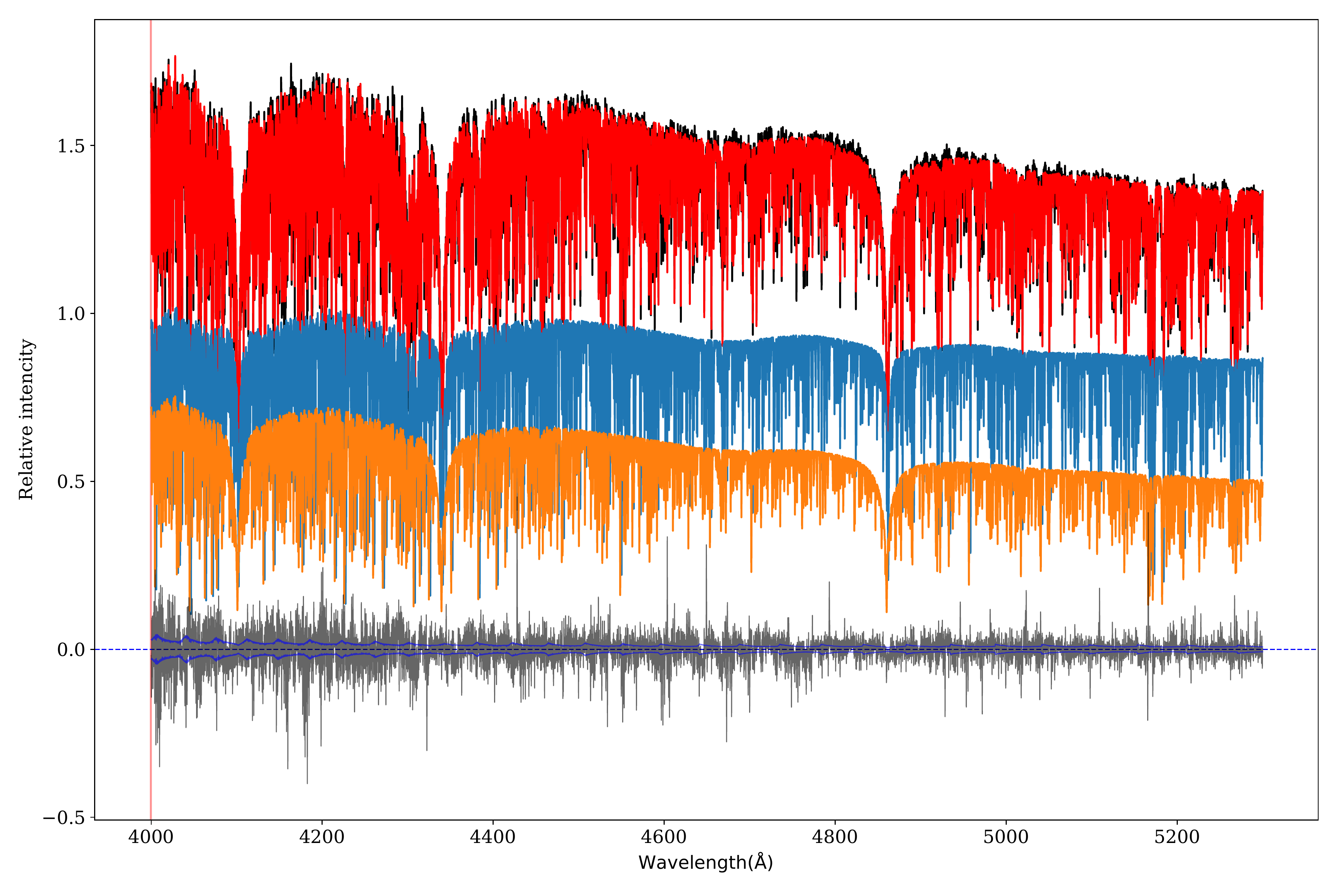}
 \includegraphics[clip=,angle=0,width=12.9cm]{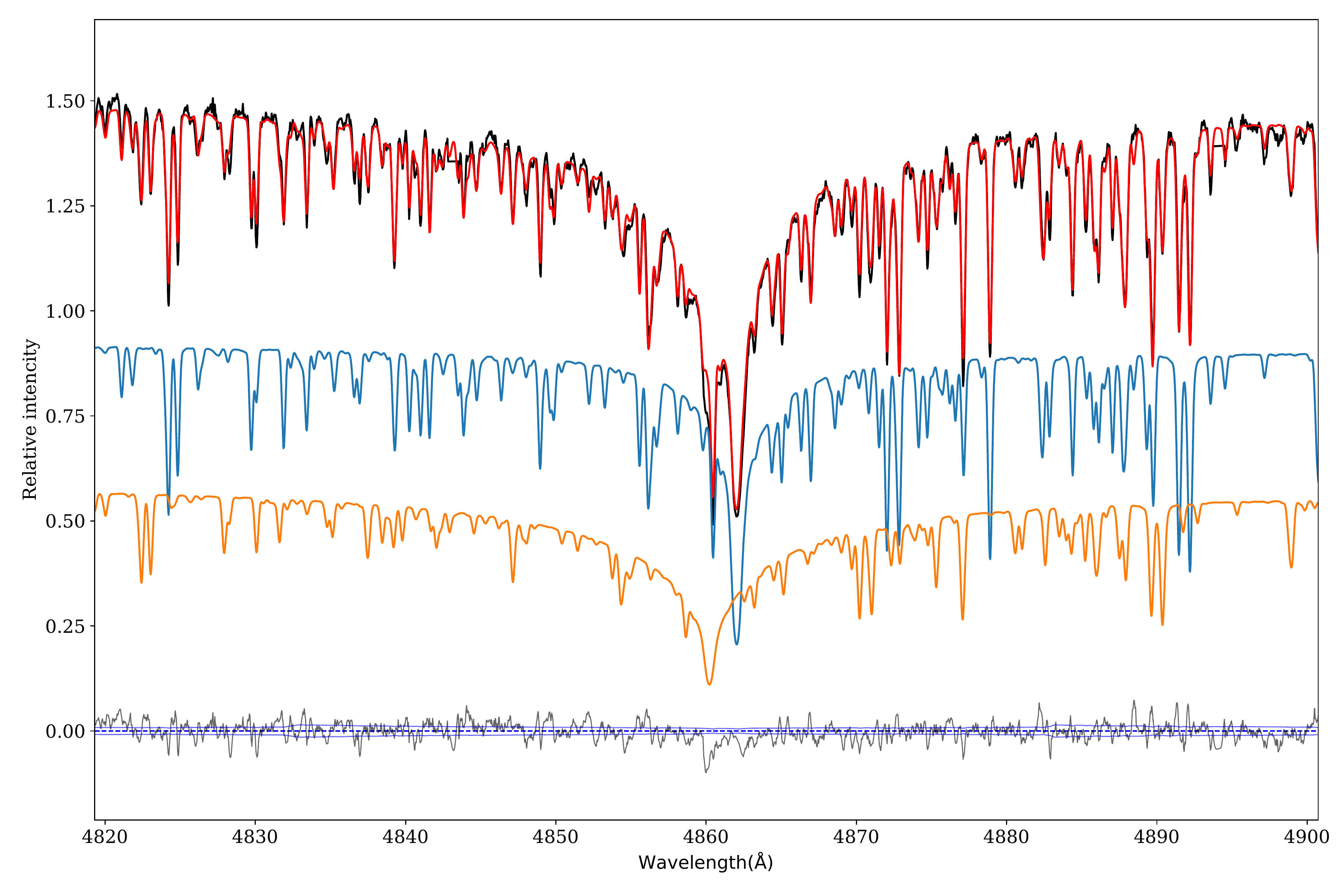}
}
 \caption{The results of the analysis of one spectrum of NN\,Del obtained with HRS. 
     The upper panel shows the result of the fit in the spectral region 4000-5300~\AA, 
     and the bottom panel shows a small spectral region near the H$\beta$ line. 
     In each panel, the black and red lines correspond to the observed spectrum and its model, respectively.
     The orange and blue lines refer to the model spectra of the first and second component,
     respectively. The bottom part of each panel shows the difference between observed spectrum
     and its model with a black line, where blue lines indicate errors in the observed spectrum.
 \label{fig:NN_Del_spec_fit}}
\end{figure*}
%========================================================================

To investigate the fully processed HRS spectra we used \fbs\ (Fitting Binary Stars) 
software specially developed by our group (Katkov et al. ~2020 in preparation) 
and partially described in \citet{2020arXiv200404115K}.
This program uses a library of theoretically calculated high-resolution stellar spectra
and is designed to determine radial velocities and stellar parameters
($T_\mathrm{eff}$, $\log g$, $\mathrm{v \sin i}$, [Fe/H])
for both components of the binary system, and parameters E(B-V) and W$_{1,2}$ that describe
the reddening of both spectra and the contribution of each component to the 
observed spectrum at the wavelength of 5500~\AA\ that is very close to the effective wavelength 
for the V-filter. By definition, $\rm W_1 + W_2 = 1$.
The program simultaneously approximates the observed spectrum by a model,
which is obtained by interpolating over the grid of stellar models
and convolves it with a function that takes into account broadening caused by the rotation
of a star ($\mathrm{v \sin i}$) with a shift corresponding to a specific value
of the radial velocity at a given epoch.
In case of a binary star, the fitting routine uses two model spectra for the components,
each with its own radial velocity and stellar atmosphere parameters,
and as result the observed spectrum is decomposed into separate spectra of two components.
If there are several spectra of the binary system obtained at different epochs,
the program finds a solution, in which the parameters
($T_\mathrm{eff}$, $\log g$, $\mathrm{v \sin i}$, [Fe/H], W)$_{1,2}$ and E(B-V)
satisfy all spectra that can be fitted simultaneously,
and velocities of both components $V_{1,2}^j$ are determined for each particular epoch $j$.
The present study uses theoretical stellar models from~\cite{Coelho14}.

%========================================================================
\begin{figure*}
\centering{
 \includegraphics[clip=,angle=0,width=0.7\textwidth]{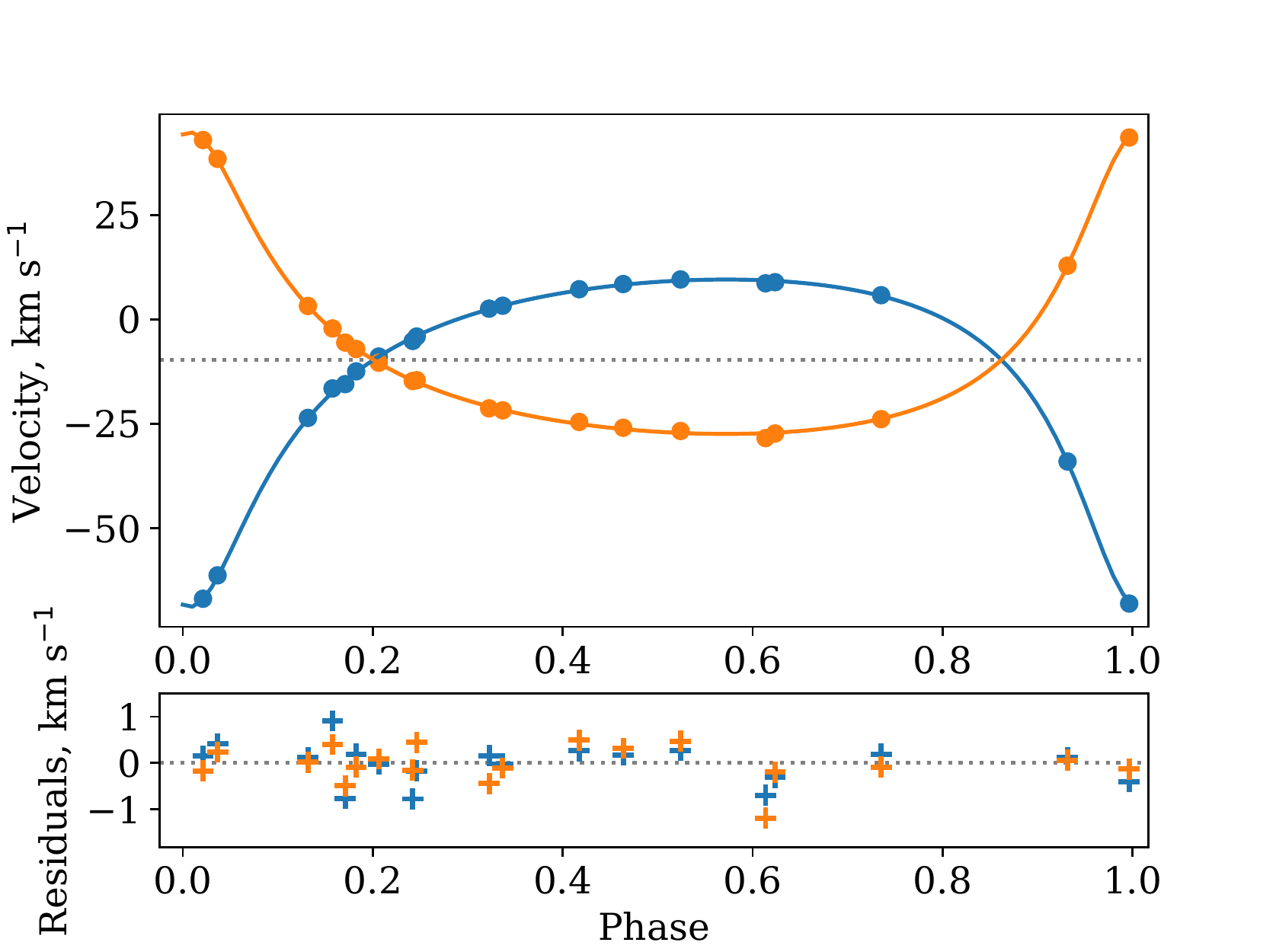}
}
 \caption{The final radial velocity curve for NN\,Del as the function of phase.
     The fitted parameters are shown in Table~\ref{tab:NN_Del_orb}.
     The model is constructed by the points listed in Table~\ref{tab:NN_Del_obs}.
     The velocity curve of component A is shown in orange, and the velocity curve of component B
     is shown in blue. The differences between the model and the observed points are shown
     in the bottom panel. 
     The final accuracy of the velocity curve (residuals after model subtraction)
     is 0.494 km s$^{-1}$, what is close to the nominal accuracy
     of 0.300 km s$^{-1}$ for HRS spectra obtained in MR mode \citep{2019AstBu..74..208K}.
 \label{fig:NN_Del_fit_Vel}}
\end{figure*}
%========================================================================

%========================================================================
\begin{figure*}
\centering{
 \includegraphics[clip=,angle=0,width=0.7\textwidth]{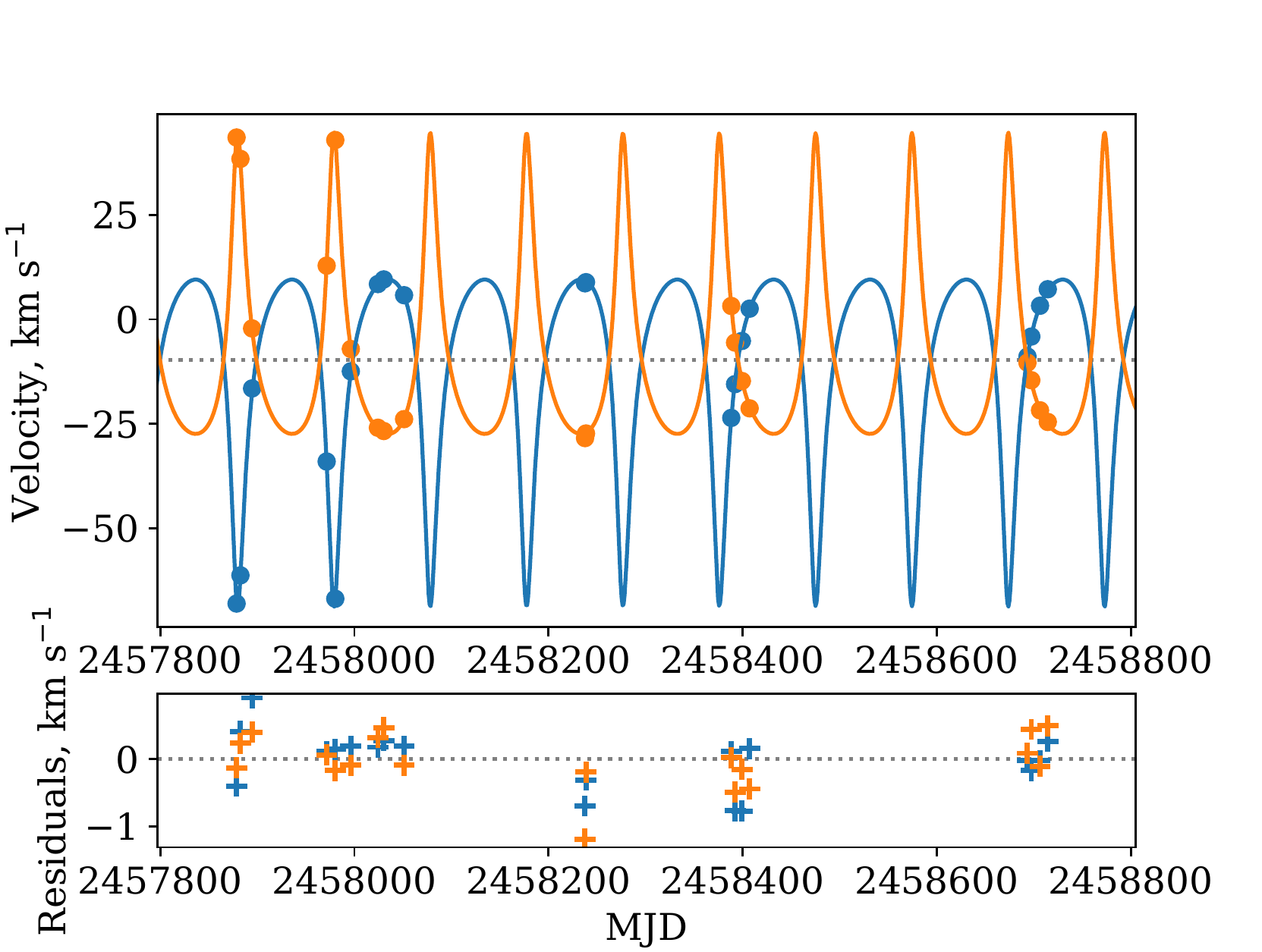}
}
 \caption{The final radial velocity curve for NN\,Del as the function of Julian date.
     Designations are the same as in Figure~\ref{fig:NN_Del_fit_Vel}.
     From this figure one can see that the observations were made during ten periods
     of NN\,Del.
 \label{fig:NN_Del_fit_Vel_all}}
\end{figure*}
%========================================================================

%**************************************************************************
\section{Results of our study of NN\,Del}
\label{txt:results}

The measured heliocentric velocities for both components of the system NN\,Del
and their errors are presented for each epoch in Table~\ref{tab:NN_Del_obs}.
To determine the velocities, as well as the parameters of the components and orbital parameters 
of the binary system NN\,Del, we used the iterative procedure described below.
The necessity of this iterative procedure was due to two facts:\\
(1) We used only spectral observations, but to obtain the complete set of parameters
to calculate masses and luminosity, 
a simultaneous analysis of spectral and photometric data is necessary, as was done, for example,
by \citet{2018MNRAS.478.1942S}. 
Potentially, masses for both components can be obtained by analyzing simultaneously 
spectral and astrometric data, as was done by \citet{2019A&A...632A..31G}, 
but for the further analysis of the evolutionary status of NN\,Del
these authors were still forced to use the parameters derived by \citet{2018MNRAS.478.1942S}.\\
(2) The presence of known temperature-$\log g$ degeneration  for stars colder 
than about 7000~K. With the full set of spectral and photometric data in hands, this 
degeneration can be avoided.

Our iterative procedure was as follows:
\begin{enumerate}
\item
During the first iteration, the program \fbs\ was used only with one limitation -- it was 
assumed that both components had the same metallicity ($[Fe/H]_1 \equiv [Fe/H]_2$;
\citep{2020MNRAS.492.1164H}).
All 19 obtained spectra were fitted simultaneously and 19 velocities were obtained
for both components together with the parameters of both components 
($T_\mathrm{eff}$, $\log g$, $\mathrm{v \sin i}$, [Fe/H])$_{1,2}$. 
The result of the analysis of a single spectrum is shown as an example
in Figure~\ref{fig:NN_Del_spec_fit}.
The upper panel of the figure shows the result of the fit in the spectral range 4000-5300~\AA,
and the bottom panel shows the result of the same fit in the H$\beta$ line region.
In each panel, the black and red lines correspond to the observed spectrum and its model.
The blue and orange lines refer to the model spectra of the first and second component,
respectively. The bottom part of each panel shows the difference between observed spectrum
and its model with a black line, where blue lines indicate errors in the observed spectrum.
\item
Based on the obtained velocities and their errors, the radial velocity curves for 
both components were constructed using the program \fbs, and
the orbital parameters of the system and their errors were calculated.
Parameters such as the orbital period $P$,
velocity semi-amplitudes of both components $K1$ and $K2$,
system eccentricity $e$, system heliocentric velocity $\gamma$ and
the longitude of the periastron $\omega$ were obtained.
\item
Using previously determined $P$, $e$, $K1$ and $K2$ and the inclination angle $i$
\citep[taken from research by][]{2018MNRAS.478.1942S}, the masses of 
M$_1$ and M$_2$ of both components and their errors were determined by the equations:
\begin{equation}
\begin{split}
\label{eqn:1}
M_1  & = 1.036149 \, 10^{-7}\,(K_1 + K_2)^2\,K_2\,P\,e^{3/2}\,\sin^{-3}(i) \\
M_2  & = 1.036149 \, 10^{-7}\,(K_1 + K_2)^2\,K_1\,P\,e^{3/2}\,\sin^{-3}(i)
\end{split}
\end{equation}
Here $M_{1,2}$ is in $M_\odot$, $K_{1,2}$ is in km/s, P is in day.
\item
Next the absolute sizes of the major and minor axes of the NN\,Del system were
calculated as:
\begin{equation}
\begin{split}
\label{eqn:2}
a_1 & = 0.01976569 \, K_1 \, P \, e^{1/2} \, \sin^{-1}(i) \\
a_2 & = 0.01976569 \, K_2 \, P \, e^{1/2} \, \sin^{-1}(i)
\end{split}
\end{equation}
as well as the radii R$_1$ and R$_2$ for both components:
\begin{equation}
\begin{split}
\label{eqn:3}
R_1 = r_1 * (a_1 + a_2) \\
R_2 = r_2 * (a_1 + a_2)
\end{split}
\end{equation}
where the relative radii ($r_1/r_2$) for both components were taken from \citet{2018MNRAS.478.1942S}, 
since they used both photometric and spectral data simultaneously,
and their relative radii are more correct compare to the ones from \citet{2003Ap&SS.283..297G}. 
Here $a_{1,2}$ and $R_{1,2}$ is in $R_\odot$.
\item
Finally, $\log g_1$ and $\log g_2$ were recalculated using:
\begin{equation}
\begin{split}
\label{eqn:4}
\log g_1 = 4.438068 + \log M_1 - 2\,\log R_1 \\
\log g_2 = 4.438068 + \log M_2 - 2\,\log R_2
\end{split}
\end{equation}
Numerical coeeficients of Eqs.~(\ref{eqn:1},\ref{eqn:2},\ref{eqn:4}) are taken
from \citet{2016ApJS..227...29P} and based on the IAU~2015 Resolution~B3 on Recommended Nominal
Conversion Constants for Selected Solar and Planetary Properties \citep{2015arXiv151007674M}.

If the new values were different from the previous values by more than
0.01~dex, all step were repeated, with the new, fixed,
values for $\log g_1$ and $\log g_2$.
When the difference between the old and new $\log g_1$ and $\log g_2$
was less of 0.01~dex, the procedure was stopped and all calculated
parameters were declared as the final ones. 
The final heliocentric velocities for both components are presented 
in Table~\ref{tab:NN_Del_obs} with their errors.
\end{enumerate}

%-----------------------------------------------------------------------------------------
\begin{table}
   \centering
   \caption{Parameters for both stellar components.}
   \begin{tabular}{lrr}
      \hline\hline
		Parameter           &       Component A          & Component B     \\ \hline
$T_\mathrm{eff}$ (K)               &  6545$\pm$180       & 6190$\pm$85     \\
$\log T_\mathrm{eff}$ (K)          &  3.82$\pm$0.02      & 3.79$\pm$0.01   \\
log g (dex)                        &  4.16$\pm$0.01      & 3.92$\pm$0.01   \\
$\mathrm{v \sin i}$ (km s$^{-1}$)  &  1.15$\pm$0.13      & 0.36$\pm$0.28   \\
W        (at 5550~\AA)             &  0.378$\pm$0.009    & 0.622$\pm$0.009 \\
$\mathrm{[Fe/H]}$ (dex)            &       \MC{2}{c}{$-0.19\pm0.05$}       \\
E(B-V)                             &       \MC{2}{c}{$0.026\pm0.002$}      \\
      \hline
   \end{tabular}
   \label{tab:NN_Del_comp}
\end{table}
%-----------------------------------------------------------------------------------------

%-----------------------------------------------------------------------------------------
\begin{table*}
   \centering
   \caption{Best-fit orbital elements.}
   \begin{tabular}{llc}
      \hline\hline
		Parameter                                         &       Value          & \%    \\ \hline
Epoch at radial velocity maximum $T_0$ (d)                & $2457779.76\pm0.06$  & 0.00  \\
Orbital period $P$ (d)                                    & 99.252$\pm$0.024     & 0.02  \\
Eccentricity $e$                                          & 0.517$\pm$0.002      & 0.38  \\
Radial velocity semi-amplitude $K1$ (km s$^{-1}$)         & 39.191$\pm$0.179     & 0.46  \\
Radial velocity semi-amplitude $K2$ (km s$^{-1}$)         & 36.101$\pm$0.099     & 0.27  \\
Systemic heliocentric velocity $\gamma$ (km s$^{-1}$)     &$-$9.692$\pm$0.066    & 0.68  \\
The longitude of the periastron $\omega$  (degrees)       & 170.307$\pm$0.532    & 0.31  \\
Root-mean-square residuals of Keplerian fit (km s$^{-1}$) & 0.406                & --    \\
      \hline
   \end{tabular}
   \label{tab:NN_Del_orb}
\end{table*}
%-----------------------------------------------------------------------------------------

%-----------------------------------------------------------------------------------------
\begin{table}[t]
   \centering
   \caption{The absolute parameters for NN\,Del}
   \begin{tabular}{lrr}
      \hline\hline
		Parameter                     &     Value          &     \%      \\ \hline
$M_1$ ($M_\odot$)                     &   1.320$\pm$0.011  &     0.83    \\
$M_2$ ($M_\odot$)                     &   1.433$\pm$0.015  &     1.06    \\
log $M_1$                             &   0.121$\pm$0.004  &     --      \\ 
log $M_2$                             &   0.156$\pm$0.005  &     --      \\ \hline
$a_1$ ($R_\odot$)                     &  65.813$\pm$0.315  &     0.48    \\
$a_2$ ($R_\odot$)                     &  60.624$\pm$0.188  &     0.31    \\
$R_1\mathrm{(spec)}$ ($R_\odot$)      &   1.594$\pm$0.016  &     1.00    \\
$R_2\mathrm{(spec)}$ ($R_\odot$)      &   2.176$\pm$0.018  &     0.82    \\
log $R_1\mathrm{(spec)}$              &   0.203$\pm$0.004  &     --      \\
log $R_2\mathrm{(spec)}$              &   0.338$\pm$0.004  &     --      \\
$R_1\mathrm{(phot)}$ ($R_\odot$)      &   1.614$\pm$0.070  &     4.35    \\
$R_2\mathrm{(phot)}$ ($R_\odot$)      &   2.314$\pm$0.071  &     3.06    \\
log $R_1\mathrm{(phot)}$              &   0.208$\pm$0.019  &     --      \\
log $R_2\mathrm{(phot)}$              &   0.364$\pm$0.013  &     --      \\ \hline
%%%%%%%%%%%%%%%%%%%%%%%%%%%%%%%%%%%%%%%%%%%%%%%%%%%%%%%%%%%%%%%%%%%%%%%%%%%%%%%%%%%
$L_1\mathrm{(spec)}$ ($L_\odot$)      &   4.164$\pm$0.346  &     8.31    \\
$L_2\mathrm{(spec)}$ ($L_\odot$)      &   6.221$\pm$0.371  &     5.96    \\
log $L_1\mathrm{(spec)}$              &   0.620$\pm$0.036  &     --      \\
log $L_2\mathrm{(spec)}$              &   0.794$\pm$0.026  &     --      \\
$L_1\mathrm{(phot)}$ ($L_\odot$)      &   4.280$\pm$0.151  &     3.53    \\
$L_2\mathrm{(phot)}$ ($L_\odot$)      &   7.051$\pm$0.205  &     2.90    \\
log $L_1\mathrm{(phot)}$              &   0.631$\pm$0.015  &     --      \\ 
log $L_2\mathrm{(phot)}$              &   0.848$\pm$0.013  &     --      \\ \hline
Distance (pc)                         &   169.71$\pm$4.27  &    2.52     \\
Age (Gyr)                             &   2.25$\pm$0.19    &    8.44     \\
Spectral Type (A)                     &             F5     &     --      \\
Spectral Type (B)                     &             F8     &     --      \\
      \hline
   \end{tabular}
   \label{tab:NN_Del_glob}
\end{table}
%-----------------------------------------------------------------------------------------

Since the output errors of the \fbs\ program are fitting errors, which are often underestimated, 
errors of the output parameters need to be estimated in more realistic way.
A statistical evaluation was used, where all possible pairs of the observed spectra were taken,
parameters $\log g_{1,2}$ were fixed and these pairs were fitted by the program. 
With 19 observed spectra 171 different pairs were fitted. 
After that, for each solution, the parameter $q$ (mass ratio of the components) 
was calculated as:
\begin{equation}
q  = -(V_{1} - \gamma)/(V_{2} - \gamma),
\end{equation}
where $\gamma$ is the systemic heliocentric velocity of the NN\,Del system.

Unstable solutions far from $q = 1$ were discarded (approximately 30\% of decisions) 
iteratively with a clipping level of $2.5\sigma$, and all other solutions were used 
to calculate the average and its error for each of the parameters
($T_\mathrm{eff}$, $\mathrm{v \sin i}$, [Fe/H])$_{1,2}$ and E(B-V).
The final calculated errors are given in Table~\ref{tab:NN_Del_comp}.
The calculated radial velocity curves as a function of phase and Julian date 
are shown in Figures~\ref{fig:NN_Del_fit_Vel} and \ref{fig:NN_Del_fit_Vel_all}, respectively.
The final accuracy of the velocity curve (scatter of the points after model 
subtraction) was 0.406~km~sec$^{-1}$, which is close to the nominal accuracy of
0.300~km~s$^{-1}$ for HRS spectra obtained in MR mode \citep{2019AstBu..74..208K}. 
The values for the orbital parameters of the binary star NN\,Del and their errors 
are given in Table~\ref{tab:NN_Del_orb}.

With all the values of the physical parameters of the NN\,Del system and their errors, 
as well as the orbital parameters of the system, the masses and radii of both components 
were calculated by Eq.\ref{eqn:1}--\ref{eqn:3}.
Using the obtained radii and temperatures, the luminosity of each of the components 
was then calculated as:
\begin{equation}
\label{eq:L}
L\mathrm{(spec)} = R^2 \left(\frac{T_\mathrm{eff}}{5780}\right)^4,
\end{equation}
where T$_\mathrm{eff}$ of the Sun is 5780~K.
The obtained absolute parameters of the NN\,Del system are presented in Table~\ref{tab:NN_Del_glob}.

%========================================================================
\begin{figure}[h]
\centering{
 \includegraphics[clip=,angle=-90, width=8.0cm]{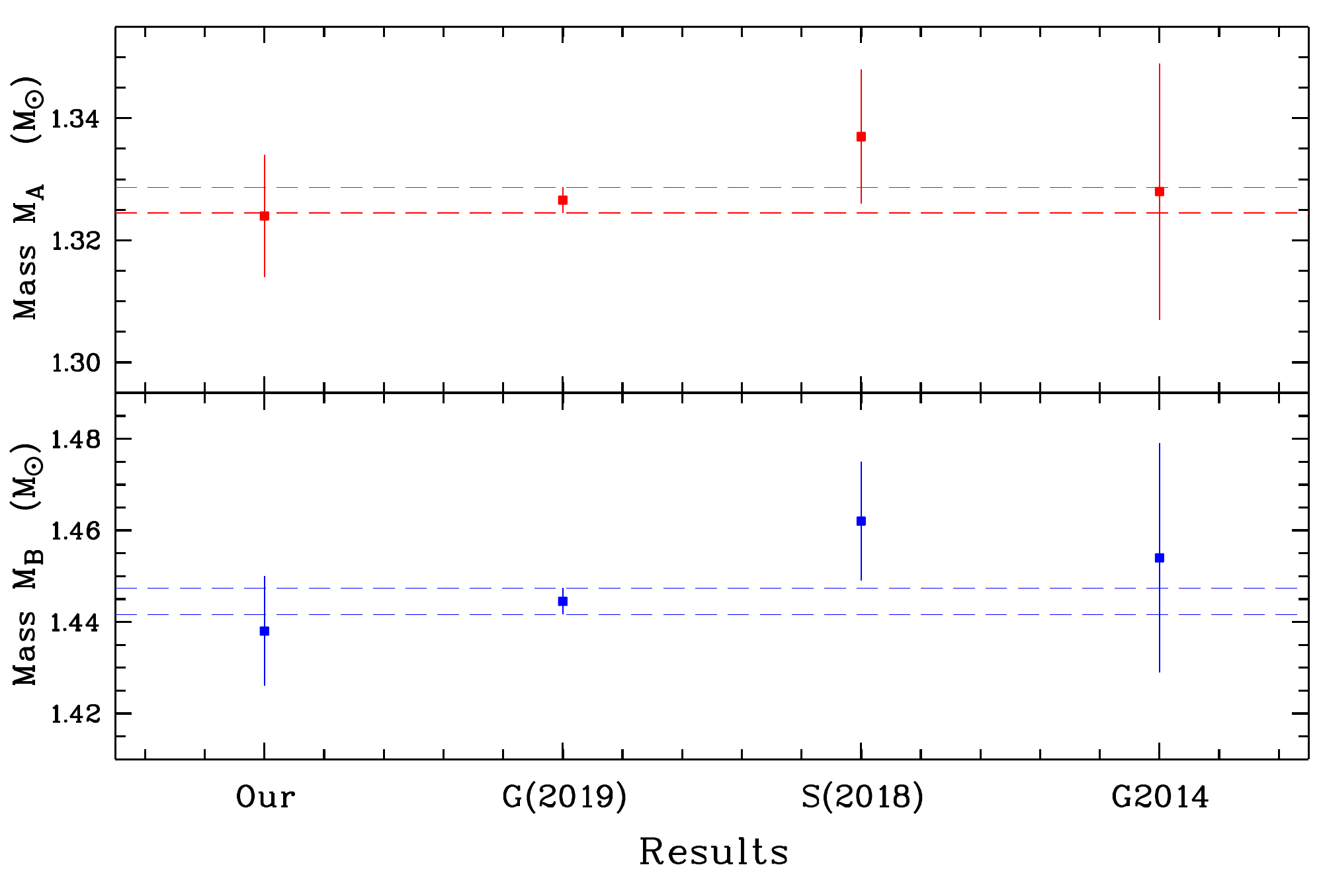}
}
 \caption{Comparison of our mass measurement for both components of NN\,Del with previously
     published results (see Table~\ref{tab:NN_Del_publ} for more details).
 \label{fig:NN_Del_comp_M}}
\end{figure}
%========================================================================

%**************************************************************************
\section{Discussion}
\label{txt:discussion}

%%%%%%%%%%%%%%%%%%%%%%%%%%%%%%%%%%%%%%%%%%%%%%%%%%%%%%%%%%%%%%%%%%%%%%%%%%%
\subsection{Comparison of parameters for the NN\,Del system}
\label{txt:dis_comp}

To evaluate the correctness of our orbital parameters calculation program,
we fitted the 40 radial velocities of the system NN\,Del given in Table~A.2
of \citet{2019A&A...632A..31G} and obtained full agreement with the orbital elements presented
in \citet{2019A&A...632A..31G} up to the final accuracy of the velocity
curves: 0.10~km~sec$^{-1}$ for the component A and 0.18~km~sec$^{-1}$ for the component B.

It was obvious from the beginning, that since we use \'echelle spectra with
over twice lower resolution compare of the CHIRON \'echelle spectrograph (R~$\sim 80000$) 
used by \citet{2018MNRAS.478.1942S} and about one and a half times lowere than the resolution 
of the STELLA \'echelle spectrograph (R~$\sim 55000$) used by \citet{2019A&A...632A..31G},
our accuracy of the obtained parameters could be worse or comparable to the accuracy 
obtained by those authors. Figure~\ref{fig:NN_Del_comp_M} shows a comparison of our mass estimates 
for both components of NN\,Del with previously published results.
It can be seen that our mass estimates do not show any systematics
and are in very good agreement with the most accurate values obtained 
by \citet{2019A&A...632A..31G}, with differences not exceeding $1\sigma$ of the total error.
That also means that the radial velocity semi-amplitudes determined by us do not show 
any systematics and positively characterize the quality of HRS data.

With found absolute parameters of the system NN\,Del, we can try to check their 
correctness. First, using the distance to NN\,Del according to $Gaia$ DR2
\citep[the parallax $5.6393\pm0.0636$ is translated to a distance $177.33\pm2.00$ parsec;][]{2018A&A...616A..11G}
as well as the color excess $E(B-V)$ resulting from our study, we can estimate the absolute 
magnitude of NN\,Del as $\mathrm{M(V)} = 2.08\pm0.03$~mag. Since we know from our spectral data
the contributions (parameter W in Table~\ref{tab:NN_Del_comp}) of each component 
at the equivalent wavelength of the V-band, we can separate the resulting absolute magnitude
into contributions of each component. Taking into account the bolometric corrections 
from \citep{1992msp..book.....S} for the $T_\mathrm{eff}$, presented in Table~\ref{tab:NN_Del_comp},
we are then able to calculate their luminosities (photometric luminosities hereafter; $L\mathrm{_{1,2}(phot)}$)
and compare these luminosities with the spectral luminosities ($L\mathrm{_{1,2}(spec)}$) 
obtained by Eq.\ref{eq:L}. 
We obtain $L\mathrm{_{2}(phot)} = 7.05\pm0.20 L_\odot$ and 
$L\mathrm{_{2}(phot)} = 4.28\pm0.15 L_\odot$, which agree with the luminosities 
$L\mathrm{_{1,2}(spec)}$ presented in Table~\ref{tab:NN_Del_glob}.
From these photometric luminosities it is possible to calculate photometric radii using
Eq.\ref{eq:L} and temperatures of the components from Table~\ref{tab:NN_Del_comp}
as $R\mathrm{_{2}(phot)} = 2.31\pm0.07 R_\odot$ and $R\mathrm{_{1}(phot)} = 1.61\pm0.07 R_\odot$.
As can be seen from the comparison of luminosities and radii, given in Table~\ref{tab:NN_Del_glob},
both spectral and photometric luminosities and radii of the component~A 
agree with a difference of only 0.3$\sigma$ and 0.5$\sigma$.
The spectral and photometric luminosities and radii of the component~B are slightly 
less agreeable at about 2.0$\sigma$ and 2.2$\sigma$, respectively, but are still
close to each other.
Such a comparison of absolute parameters, counted in a different way and matched to each other,
is a good indicator of the correctness of the absolute parameters found for the NN\,Del system.

%========================================================================
\begin{figure}[t]
\centering{
 \includegraphics[clip=,angle=-90, width=8.0cm]{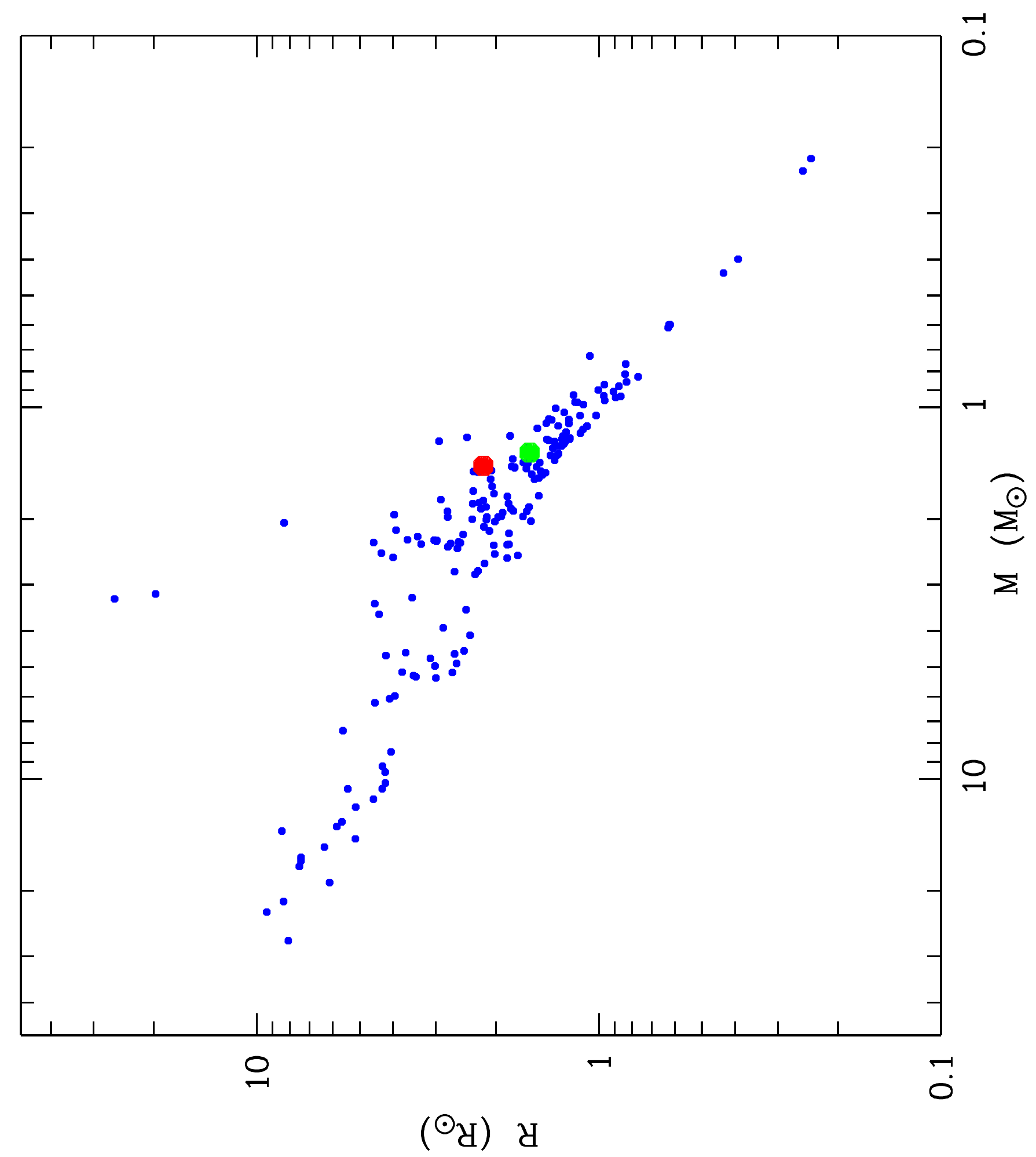}
 \includegraphics[clip=,angle=-90, width=8.0cm]{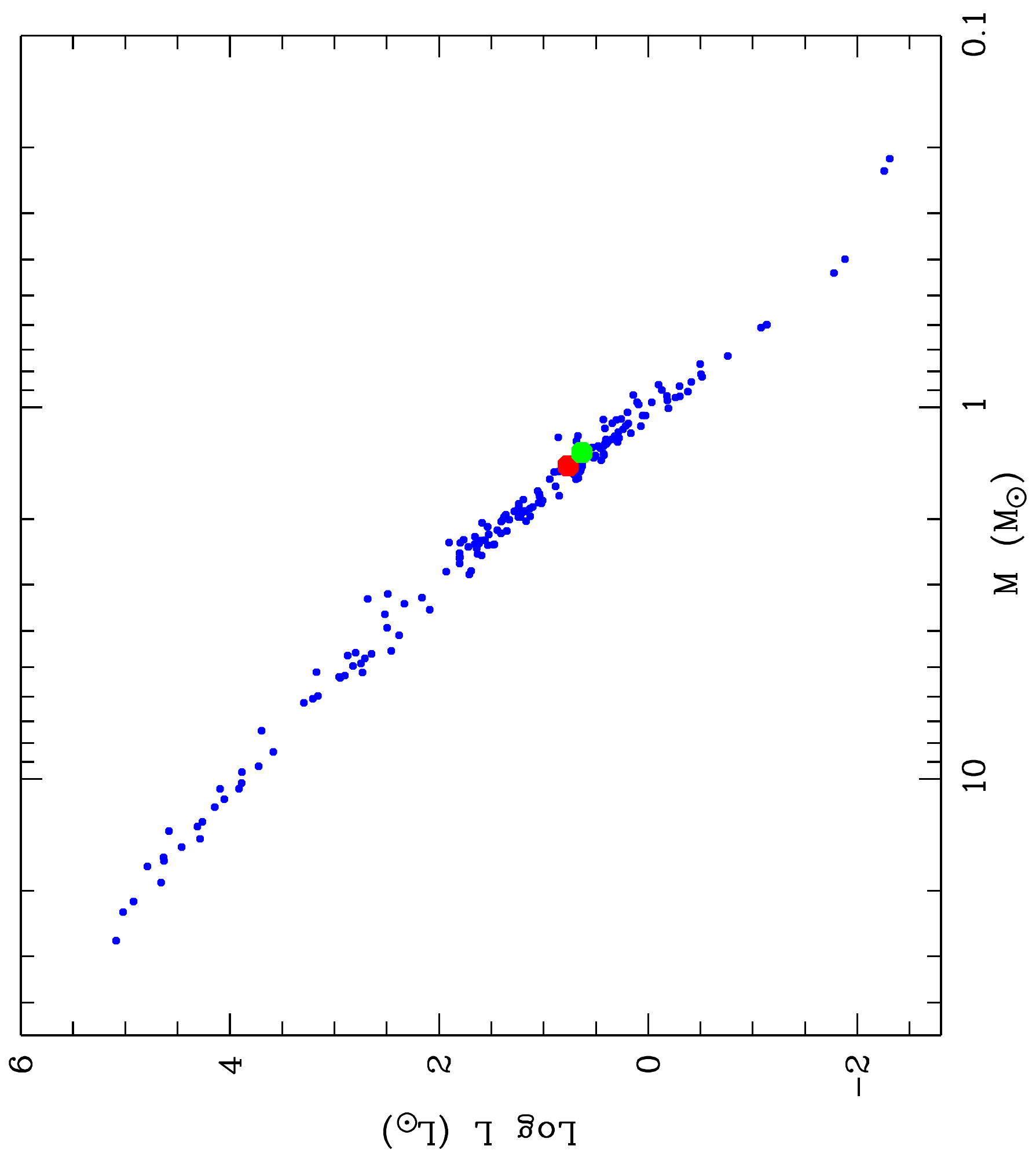}
}
 \caption{Diagram for all stars from \citet{2010A&ARv..18...67T} with position of both
     components of NN\,Del. Component A shown in green and component B in red.
     Sizes of the symbols are significantly larger than the errors.
\label{fig:NN_Del_MLR}}
\end{figure}
%========================================================================

We can also independently estimate the distance to NN\,Del and compare it with the distance 
from $Gaia$ DR2 \citep{2018A&A...616A..11G}. 
Our estimation of the distance could be done with use of the known total V$_\mathrm{tot}$ magnitude,
found extinction shown in Table~\ref{tab:NN_Del_comp} and an absolute magnitude 
M(V$_\mathrm{tot}$). The M(V$_\mathrm{tot}$) could be found as the sum of absolute magnitudes
for both components M(V$_\mathrm{1,2}$), where the absolute magnitude for each component could be
calculated with use of luminosities based on the spectroscopic radii $L_\mathrm{(spec)}$ 
from Table~\ref{tab:NN_Del_glob} and temperatures $T_\mathrm{eff}$ from Table~\ref{tab:NN_Del_comp} 
and bolometric corrections calculated with use of \citep{1992msp..book.....S}.
The final result is also shown in Table~\ref{tab:NN_Del_glob}. 
Our estimated distance is smaller than the distance according to $Gaia$ 
and this result is in the agreement with the result obtained by \citet{2019A&A...632A..31G}
as $\mathrm d=167.99\pm0.65$~pc.
However, we have to admit here that the large error of our assessment means that
difference we found is not significant with a difference of only 1.62$\sigma$.
%Our analysis shows that the largest contribution 
%to our error comes from the reddening error $E(B-V)$.

In the \citet{2010A&ARv..18...67T} review the data on luminosities and masses of 190 stars, 
from 95 DLEB systems, have been collected. A comparison of our obtained characteristics 
for NN\,Del with data from \citet{2010A&ARv..18...67T}, as presented in Figure~\ref{fig:NN_Del_MLR},
shows that our characteristics of the NN\,Del components are fully consistent with the
masses, luminosities and radii of the stars in this large sample.

%========================================================================
\begin{figure}
\centering{
 \includegraphics[clip=,angle=-90, width=8.0cm]{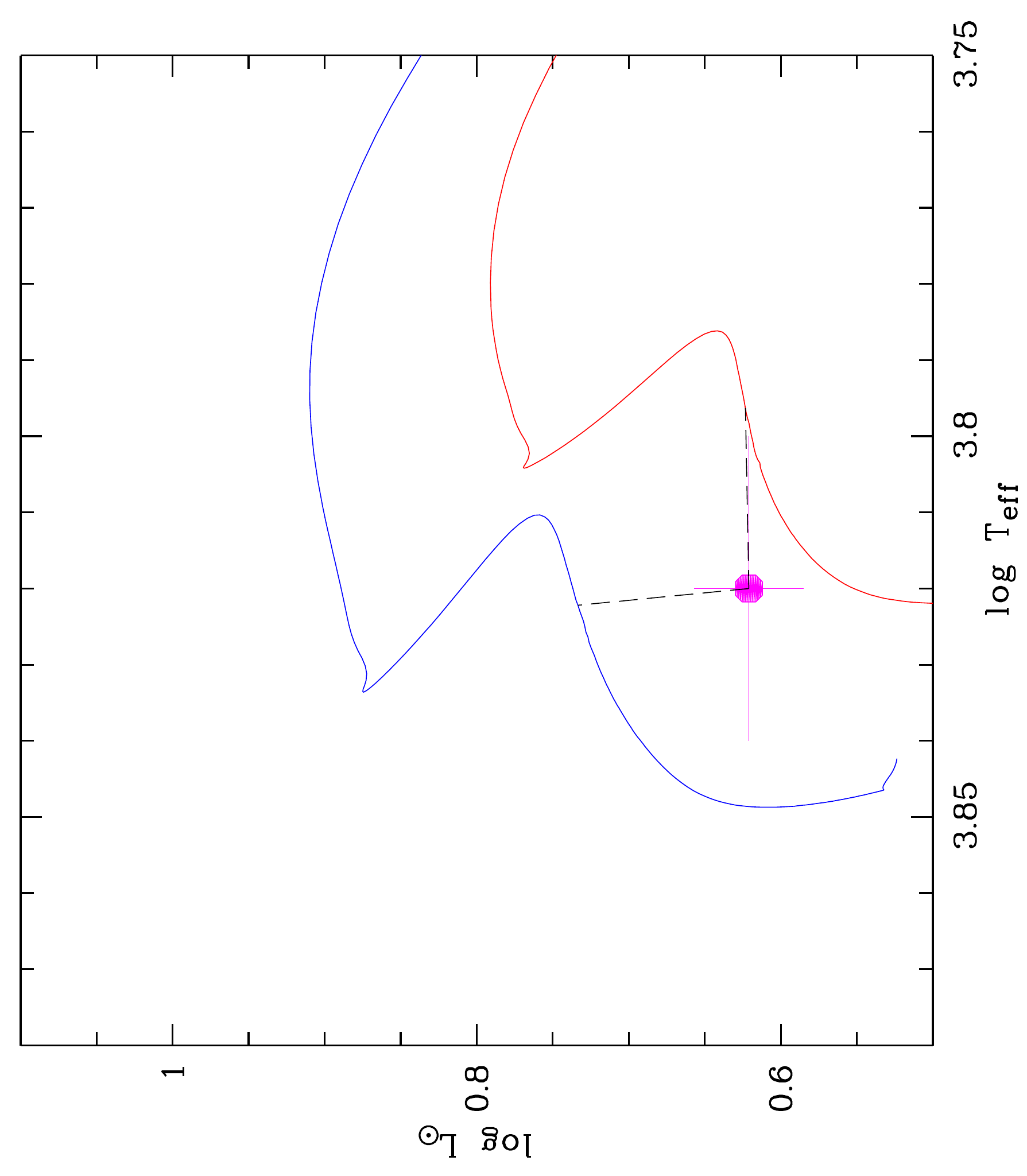}
 \includegraphics[clip=,angle=-90, width=8.0cm]{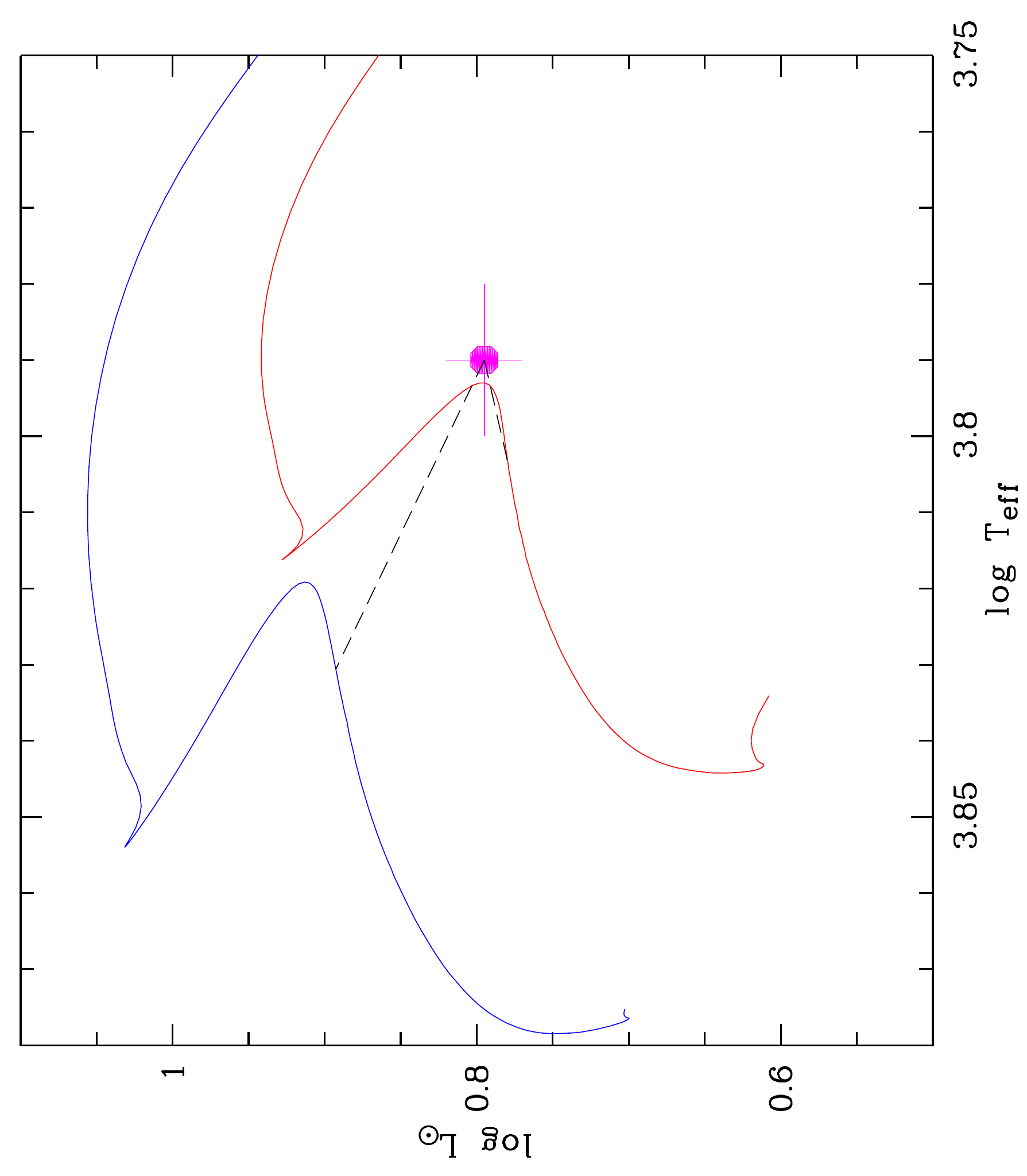}
}
 \caption{Evolutionary tracks of stars of mass 1.44~$M_\odot$ (upper panel) for the component A 
     and mass 1.32~$M_\odot$ (lower panel) for the component B for two metallicities:
     [Fe/H] = -0.19~dex (blue) and [Fe/H] = 0.05~dex (red).
     The burgundy dots include the NN\,Del system components,
     and the black dotted lines show the distance to the optimal solution as
     found by Eq.\ref{eq:fit}.
\label{fig:NN_Del_evol}}
\end{figure}
%========================================================================

%%%%%%%%%%%%%%%%%%%%%%%%%%%%%%%%%%%%%%%%%%%%%%%%%%%%%%%%%%%%%%%%%%%%%%%%%%%
\subsection{Evolutionary state}
\label{txt:dis_evol}

Since NN\,Del is a detached system where components do not influence each other's evolution, 
it is reasonable to assume that evolutionary status 
of each component can be assessed based on evolutionary models of single stars. 
Since \citet{2019A&A...632A..31G} showed that results from the models \parsec\ \citep[PAdova and TRieste Stellar Evolution Code,][]{2012MNRAS.427..127B}, 
\basti\ \citep[Bag of Stellar Tracks and Isochrones,][]{2004ApJ...612..168P} and
\mist\ \citep[MESA Isochrones and Stellar Tracks,][]{2016ApJ...823..102C} agree well, 
we only used the \mist\ model in our work.
Since the masses of the NN\,Del components are known to us with the highest degree of accuracy, 
we retrieved from \mist\footnote{\url{http://waps.cfa.harvard.edu/MIST/interp_isos.html}}
the evolutionary tracks for stars of these masses and considered the positions 
of each of the components on these tracks, varying the metallicity [Fe/H] from -0.2 to 0.2~dex 
in increments of 0.05~dex. We then searched for the minimum of the equation:

{\footnotesize
    \begin{equation}
\label{eq:fit}
\chi^2 = \sum_{i=1}^2 \left[ \left(\dfrac{\Delta L}{\sigma_L}\right)_i^2 + 
\left(\dfrac{\Delta T_\mathrm{eff}}{\sigma_\mathrm{T_{eff}}}\right)_i^2 +
\left(\dfrac{\Delta R}{\sigma_R}\right)_i^2 +
\left(\dfrac{\Delta g}{\sigma_{g}}\right)_i^2 \right],
\end{equation}
}
where the summation is done on both components ($i = 1,2$), the symbol $\Delta$ 
shows the logarithmic difference between the model and the value obtained from the observations, 
and $\sigma$ is also used in the logarithmic scale. 
The search was performed for each model metallicity under the assumption that both components 
of the system have the same metallicity.

Figure~\ref{fig:NN_Del_evol} shows the $\log L$--$\log \mathrm{T_{eff}}$ plot for two
solutions: (1) for metallicity [Fe/H] = $-0.19$~dex, 
which is derived from our star models (see Section~\ref{txt:results})
and (2) for metallicity [Fe/H] = 0.05~dex, which showed the best $\chi^2$ from Eq.\ref{eq:fit}.
For metallicity [Fe/H] = $-0.19$~dex, the age of the system NN\,Del is estimated to be 2.06~Gyr,
and for [Fe/H] = 0.05~dex it is 2.44~Gyr.
In both cases, as can be seen in Figure~\ref{fig:NN_Del_evol},
both components have not yet passed the turn point and are on the main sequence,
which coincides with the conclusion by \citet{2019A&A...632A..31G}.

It is possibly to explain the resulting difference in metallicities by the systematic shift
between the metallicity scales of the stellar models from \citet{Coelho14} and the \mist\ models.
We therefore took stellar models from \textsc{phoenix} \citet{2013A&A...553A...6H} 
and recalculated our data according to the methodology described
in Sections~\ref{txt:spec_analysis} and \ref{txt:results},
and resulted the value [Fe/H] = $-0.20$~dex as well.
We would like to note that the \textsc{phoenix} models match the observed spectra
less well and therefore we prefer to work with the models from \citet{Coelho14}.
As the final result, we adopt the age assessment for the NN\,Del system as 
$t_\mathrm{avg} = 2.25\pm0.19$~Gyr, which also agrees with
the output from \citet{2019A&A...632A..31G}.

Since \fbs\ software splits observational spectrum into two separate components
(see Figure~\ref{fig:NN_Del_spec_fit}) and we know that both of them have luminosity
class V, we can estimate the spectral type of each of them. 
For both components  Ca\,K/(H$\epsilon$+Ca\,H) intensity ratio $\sim$1,
but the CH G-band at $\sim$4300~\AA\ is not clear visible yet implies that both spectra
have spectral types F0--F5 \citep{2003MNRAS.345.1223E,2004MNRAS.353..601E}.
Additionally, taking into account criteria from \citet{1990clst.book.....J} about
EWs of different lines, I finally classify the component A as F5 based on
EW(\ion{Fe}{i}~$\lambda$4045)$=0.51\pm0.02$~\AA,
EW(\ion{Sr}{ii}~$\lambda$4077)$=0.31\pm0.02$~\AA,
EW(H$\gamma$)$=4.95\pm0.10$~\AA\ and
EW(H$\beta$)$=5.15\pm0.10$~\AA\ and component B as F8 based on
EW(\ion{Fe}{i}~$\lambda$4045)$=0.71\pm0.01$~\AA,
EW(\ion{Ca}{i}~$\lambda$4226)$=0.75\pm0.02$~\AA,
EW(H$\gamma$)$=4.20\pm0.10$~\AA\ and
EW(H$\beta$)$=4.05\pm0.10$~\AA.
This classification for component A coincides with the classification from 
\citet{2003Ap&SS.283..297G} and for components B with 
the classification from \citet{2014Obs...134..109G}.

%**************************************************************************
\section{Conclusions}
\label{txt:summ}

The long-period eclipsing binary star NN\,Del has been studied
using spectral data obtained with the HRS \'echelle spectrograph of the SALT telescope.
A velocity curve has been constructed which is based on 19 spectra obtained during
2017--2019 years and covers all phases of this binary system.
The obtained orbital and absolute parameters
of the components, together with their errors, agree with the results obtained in early works,
which shows the good quality of HRS spectra and correctness of the applied processing and 
analysis methodology. The luminosities of both components are calculated
and the effective temperatures are directly evaluated
altogether with the metallicity of the system. The analysis of the calculated parameters
is presented that allows to estimate the obtained absolute parameters
in various ways and shows the repeatability of these assessments.
The age and evolutionary status of both components have been estimated,
as well as their spectral types.

%**************************************************************************
\section*{Acknowledgments}

A.\,K. is grateful to D.\,Graczyk (the referee) for useful comments and suggestions on the manuscript. 
This work is based on observations obtained with the Southern African Large Telescope (SALT)
under programs  \mbox{2017-1-MLT-001} and \mbox{2019-1-SCI-004} (PI: Kniazev).
This research is supported by the National Research Foundation (NRF) of South Africa
and the Russian Foundation for Basic Research grant 19-02-00779.

\bibliographystyle{spr-mp-nameyear-cnd}
\bibliography{NN_Del_SALT}

\end{document}